# In Search of Subsurface Oceans within the Uranian Moons

**C. J. Cochrane[1], S. D. Vance[1], T. A. Nordheim[1], M. Styczinski[2], A. Masters[3], L. H. Regoli[4]**

[1]Jet Propulsion Laboratory, California Institute of Technology

[2]University of Washington

[3]Imperial College London

[4]John Hopkins University, Applied Physics Laboratory

Corresponding author: Corey J. Cochrane ([corey.j.cochrane@jpl.nasa.gov](corey.j.cochrane@jpl.nasa.gov))

**Key Points:**

- Favorable geometry of the Uranus system facilitates magnetic induction investigation of potential oceans within the primary moons
- Sub-surface ocean detection and constrained characterization is possible within these moons, even in the presence of an ionosphere

## Abstract

The *Galileo* mission to Jupiter discovered magnetic signatures associated with hidden sub-surface oceans at the moons Europa and Callisto using the phenomenon of magnetic induction. These induced magnetic fields originate from electrically conductive layers within the moons and are driven by Jupiter's strong time-varying magnetic field. The ice giants and their moons are also ideal laboratories for magnetic induction studies. Both Uranus and Neptune have a strongly tilted magnetic axis with respect to their spin axis, creating a dynamic and strongly variable magnetic field environment at the orbits of their major moons. Although *Voyager-2* visited the ice giants in the 1980s, it did not pass close enough to any of the moons to detect magnetic induction signatures. However, *Voyager-2* revealed that some of these moons exhibit surface features that hint at recent geologically activity, possibly associated with sub-surface oceans. Future missions to the ice giants may therefore be capable of discovering sub-surface oceans, thereby adding to the family of known "ocean worlds" in our solar system. Here, we assess magnetic induction as a technique for investigating sub-surface oceans within the major moons of Uranus. Furthermore, we establish the ability to distinguish induction responses created by different interior characteristics that tie into the induction response: ocean thickness, conductivity, and depth, and ionospheric conductance. The results reported here demonstrate the possibility of single-pass ocean detection and constrained characterization within the moons of Miranda, Ariel, and Umbriel, and provide guidance for magnetometer selection and trajectory design for future missions to Uranus.

## Plain Language Summary

The best evidence for the existence of an ocean within Jupiter's moons of Europa and Callisto comes from interpretations of measurements collected from the magnetometer on the Galileo spacecraft. Their salty oceans are able to conduct electrical currents that are driven by Jupiter's strong time varying magnetic field, producing a unique secondary magnetic field signature that can be measured externally by as spacecraft passing by. Some of the large moons of the Ice Giant planets of Uranus and Neptune are also thought to possibly contain subsurface oceans. Unfortunately, Voyager 2 did not pass close enough to any of these planetary bodies for its onboard magnetometer to sense magnetic field signatures associated with a sub-surface ocean. In this work, we investigate the feasibility of detecting magnetic signatures of sub-surface oceans within the moons of Uranus based upon various assumed interior structure profiles, thus providing guidance to required magnetometer performance and trajectory design for possible future missions.

## 1 Introduction

Uranus possesses a rich and diverse satellite system, including the major moons Miranda, Ariel, Umbriel, Titania, and Oberon, as well as host of smaller regular and irregular satellites (22 discovered to date). This contrasts with Neptune, whose only major moon, Triton, is likely a captured Kuiper Belt Object that catastrophically disrupted Neptune's original satellite system (Agnor and Hamilton, 2006). Thus, Uranus presents us with what is likely our best opportunity for in-situ study of a native ice giant satellite system. The *Voyager 2* flyby of Uranus in 1986 provided the first, and so far only, detailed imaging of the Uranian moons (e.g. Schenk et al., 2020). These images revealed an unexpectedly diverse range of surface geologies, hinting at the possibility that these objects, at least at some point in their recent pasts, may have experienced significant heating (e.g. Croft and Soderblom, 1991).

In particular, the innermost major moons Miranda and Ariel show evidence of widespread tectonism in the form of rift canyons and extensive faulting (Beddingfield et al., 2015; Beddingfield and Cartwright, 2020; Croft and Soderblom, 1991; Peterson et al., 2015; Smith et al., 1986). The surface of Miranda contains large banded regions known as coronae that may be surface expressions of extension above buoyant diapirs within the moon's interior (Pappalardo et al., 1997). The surface of Ariel appears to have overturned relatively recently, and displays extensive rift-like tectonic features (Croft and Soderblom, 1991; Schenk et al., 2020; Zahnle et al., 2003). Landforms on the surfaces of both Miranda and Ariel have been interpreted as cryovolcanic in origin (Schenk, 1991), possibly indicating that material from the interior has been emplaced onto the surface of these moons. In addition, ground-based spectroscopic observations have revealed the presence of ammonia-bearing species on the surface of Ariel (Cartwright et al., 2020, 2018). These species are expected to be destroyed by magnetospheric charged particles over relatively short timescales (Moore et al., 2007), and their presence on the surface of Ariel is therefore consistent with geologically recent emplacement (Cartwright et al., 2020). These observations are consistent with sub-surface liquid water oceans within Miranda and Ariel, perhaps even persisting to present day.

Furthermore, thermal evolution modeling results suggest that the outermost moons Titania and Oberon could have retained subsurface oceans if they contain just a few percent by total volume of ammonia (Hussmann et al., 2006). Thus, the Uranian system represents a compelling target in the search for Ocean Worlds and potentially habitable environments beyond Earth, with Miranda and Ariel highlighted as priority targets for future exploration by the recent NASA Roadmap to Ocean Worlds (Hendrix et al., 2019).

Magnetic induction had a key role in the discovery of subsurface liquid water oceans within Jupiter's moons Europa, Ganymede, and Callisto (Kivelson et al., 1999, 2002), as well as a possible magma ocean within the moon Io (Khurana et al., 2011). Because of this great utility, magnetic sounding is also a key component of the upcoming ESA *Jupiter Icy Moons Explorer* (*JUICE*, Grasset et al., 2013) and NASA's *Europa Clipper* flagship missions (Howell and Pappalardo, 2020). The *Voyager 2* flyby of Uranus revealed a uniquely complex planetary magnetic field about 50 times stronger than that of Earth's ($3.9 \times 10^{17}$ T m$^3$), tilted by 59° relative to its spin axis (also to orbital axis of the major moons), and offset from the center of the planet by ~0.3 $R_U$ (Connerney et al., 1987; Ness et al., 1986; Paty et al., 2020). Thus, like the Galilean moons, the major moons of Uranus are exposed to a strong and highly dynamic magnetic field environment (Cochrane et al. 2021, Weiss et al. 2021). However, unlike the Galilean moons, the extreme tilt of Uranus' magnetic axis causes the moons to effectively travel through a range of magnetic environments (L-shells) during one Uranus rotation period. The moons Titania and Oberon pass in and out of the magnetopause boundary, so their local magnetic field environments are expected to be even more variable than those of the inner moons.

The tilt in the magnetic dipole axis allows each of the moons to experience a time-varying field, at a frequency corresponding to the respective synodic period of Uranus. Additional time variations in the magnetic field can arise if the moons exhibit an eccentric or inclined orbit, due to the change in distance between the two bodies. All five major moons experience strong synodic variations, but do not experience strong variations at the orbital periods as each of their orbits are close to circular (i.e. low eccentricity). Miranda is the only moon with significant variation at the orbital period, as its orbital plane is inclined by 4.2°. It is also worth noting that the orbital period of Miranda is shorter than the synodic period, which is rare for moons in the Solar System.

A time-varying magnetic wave will penetrate the conducting portion of the moons according to a skin depth relationship, which is related to the frequency of the wave and conductivity of the medium. Studying multiple frequencies can therefore be advantageous for probing various depths of conducting layers within the moon (Khurana et al. 2002). These time-varying magnetic fields produce an electric field capable of inducing electric currents within each of the moons if they contain conductive layers (e.g. a saline ocean) within their interiors. These currents will induce a magnetic moment that depends on the interior properties of the moon (e.g. depth, thickness, and conductivity of internal layers) and is accompanied by a secondary dipolar magnetic field that can be detected in the vicinity of the moon using a magnetometer. At the Uranian moons, strong magnetic field variations occur at multiple frequencies, underscoring the potential utility of magnetic induction to search for sub-surface oceans within these objects.

The Uranus system offers important contrasts to other planetary systems that can help us understand how these systems evolve. The closest analog appears to be Saturn's mid-sized moons, which have undergone orbital migration that created mutual tidal heating (Neveu and Rhoden 2019). From recent discoveries there, and at distant Pluto (Nimmo et al. 2016), the prospect for oceans in some Uranian moons seems plausible. The large moons of Uranus appear to have engaged in orbital resonances that placed them in their current orbital arrangement (Ćuk et al. 2020). Thus, the moons may have experienced periods of enhanced tidal heating during their histories, a likely source for the heating needed to form the coronae on Miranda and perhaps also for prolonging the timescale over which sub-surface liquid water oceans could be sustained within the moons' interiors. (Plescia 1988, Beddingfield and Cartwright 2020).

Although models assuming simple compositions of silicates and water predict that an ocean within Miranda would have frozen out by the present era (Hussmann et al. 2006), the presence of volatile clathrates or other insulating materials could support the persistent presence of sub-surface liquid by providing thermal insulation and adding rigidity to the ice, both of which lower the Rayleigh number and thus inhibit solid-state convection (Croft 1987, Castillo-Rogez et al. 2019, Kamata et al. 2019). Within Saturn's moon Titan and the dwarf planets Ceres and Pluto, methane clathrates may impede solid state convection in ice, leading to inefficient cooling that slows the freeze-out of subsurface oceans. Similarly, the presence of ammonia on the surface of Ariel may indicate recent geological activity and suggests a role for volatiles in retaining heat and suppressing the freezing point of sub-surface liquid water layers within this moon (Cartwright et al. 2020).

In this work, we calculate the expected magnitude and frequency of the magnetic waves at the Uranian moons with the greatest potential for magnetic sounding (Miranda, Ariel, and Umbriel) and forward model their induced magnetic moment responses for thousands of different combinations of ocean thickness, ocean conductivity, ice shell thickness, and ionosphere conductance. This work demonstrates the ability to distinguish between these different ocean characteristics and provides insight into the implications that the presence of an ionosphere may have on the recoverability of ocean signatures from magnetometer data. We also discuss how the geometry of the Uranus system and the uncertainty in our knowledge of the planet's rotation rate impacts the optimal spacecraft time-of-arrival to maximize the induction signal, while also minimizing any plasma induced fields that may arise from corotating plasma currents confined to Uranus' magnetic equator.

## 2 Magnetic Field of Uranus

The strength of the induced magnetic field that can arise from the interior of the moon is bounded by the strength of the externally applied magnetic field driving induction. Here, Uranus is the source of the external field—which likely originates from a self-sustaining dynamo acting within a thin, electrically conductive, convective shell (Ruzmaikin 1991, Hubbard et al. 1995, Podolak et al., 1995, Stanely et al. 2004, Stanley et al. 2006). Of the Uranus magnetic field models available derived from the *Voyager 2* measurements (e.g., $Q_3$ Connerney1987, $AH_5$ Herbert2009), we use the $AH_5$ (Auroral Hexadecapole L = 5') which considers the available Voyager 2 magnetometer data, as well as auroral observations for improved accuracy. This model involves the standard spherical harmonic multipole expansion to represent the magnetic potential; the field is obtained by taking the negative gradient of the scalar potential, i.e. $\boldsymbol{B} = -\nabla V$. The spherical harmonic expansion for the scalar potential $V$ is expressed as

$$V = R_U \sum_{n=1}^{N} \left(\frac{R_U}{r}\right)^{n+1} \sum_{m=0}^{n} \{P_n{}^m(cos\theta)[g_n{}^m \cos(m\phi) + h_n{}^m \sin(m\phi)]\}, \quad (1)$$

where $R_U$ is equatorial radius of Uranus (taken to be 24,764 km), $r$ is the radial distance from the center of mass, $\theta$ is the colatitude, $\phi$ is the east longitude, $P_n{}^m(cos\theta)$ are the Schmidt-normalized Legendre functions of degree $n$ and order $m$, and $g_n{}^m$, $h_n{}^m$ are the internal Schmidt coefficients (where $N = 4$). Although the fields associated with higher-order quadrupole and octupole terms decrease in strength more rapidly than the dipolar terms as a function of distance from the planet, they are still able to drive magnetic induction within the moons at frequencies corresponding to the higher order harmonics of the fundamental synodic period. For this initial study, we do not include an external contribution from the solar wind and interplanetary magnetic field but note the importance of this component for assessing the magnetic field at Titania and Oberon due to the magnetopause currents at these outermost major moons. In future work, we plan to implement the effect of the solar wind in the calculation of induced magnetic fields. Here we have, however, focused on magnetic induction within the innermost major moons Miranda, Ariel, and Umbriel, which are not strongly affected by the magnetopause currents. Figure 1 illustrates the magnetic field lines of the $AH_5$ magnetic field model for an arbitrary snapshot in time on Jan 7, 2031; the color denotes the magnetic field magnitude. In the left panel, the 59° tilt of the magnetic dipole axis (magenta line)—defined by $\boldsymbol{M} = 4\pi R_U{}^3 [g_1^1, h_1^1, g_1^0]/\mu_0$— is shown with respect to the spin axis of Uranus (white line), which is tilted 97° with respect to the vector normal to the orbital plane of the planet. Note also that the magnetic axis does not pass through the center of the planet, but rather is offset by 0.3 $R_U$. As a result, the magnetic field is not the same at all locations on the surface of Uranus, but rather can vary from roughly 10,000 to 100,000 nT. The right panel of the figure, generated from ephemeris data evaluated with SPICE, illustrates the field lines from a view that is roughly normal to the orbital plane of the five moons. As illustrated, the orbits of the moons are nearly circular, resulting in little variation in the distance between each of the moons and the planet. As will be demonstrated in the next section, each moon therefore experiences only low-amplitude variations in the magnetic field at their respective orbital periods. Miranda is the only body with a significant component at the orbital period, due to its inclination of 4.2°. Despite their low eccentricities, the tilt of the Uranus dipole axis with respect to the spin axis causes large variations in the magnitude of the background magnetic field at the synodic period of each moon—320 nT at Miranda to 3.6 nT at Oberon (neglecting for the moment the field contributions from the magnetopause currents to be discussed later). Assessing the time-variable portion of these fields is important as it is strictly the oscillating portion that drives magnetic induction within conducting material.

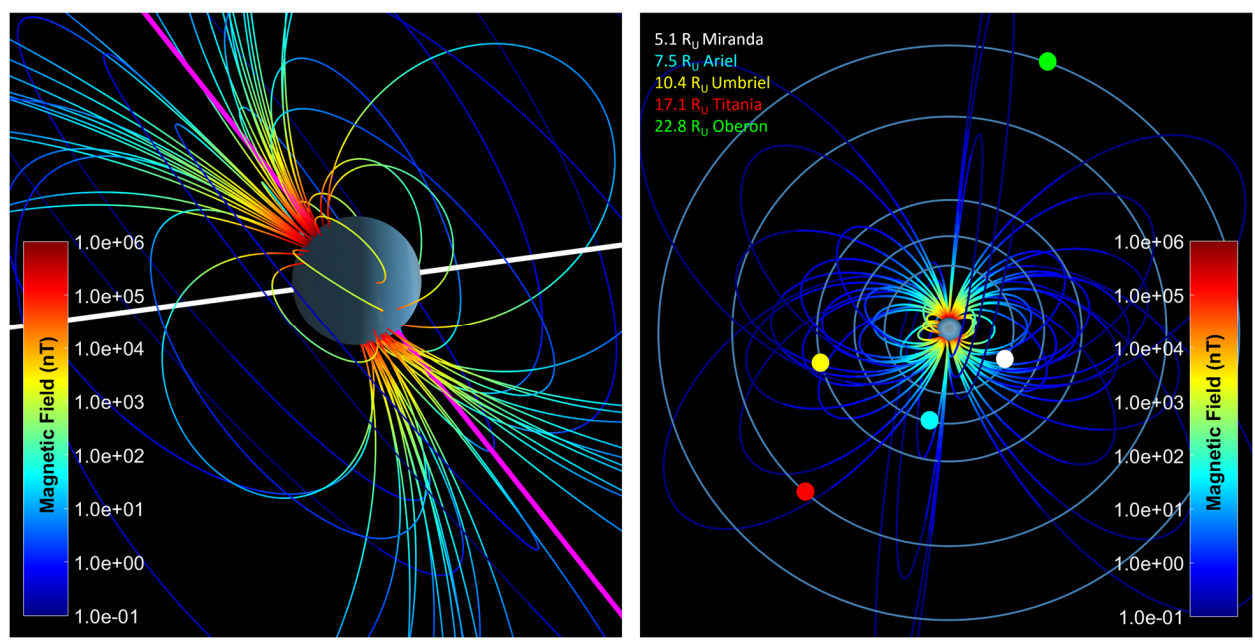

**Figure 1.** (left) Simulation of Uranus' magnetic field lines, color coded by strength, at a specific instant in time which highlights the magnetic dipole axis tilt, offset, and higher order structure. (right) A more distant view of Uranus' magnetic field lines as viewed from its North pole. Also illustrated are the moons and their orbits around the planet.

## 3 Magnetic Field at the Uranian Moons

Here we provide an initial assessment of the variability of Uranus' magnetic field at the positions of the five major moons. Although the orbital variations between the moons and planet are small, large variations in the magnetic field are observed at each of the moons due to the large tilt of the magnetic axis with respect to the spin axis of Uranus and the orbital plane of the moons. These variations do not occur at the rotation rate of the planet, but rather at the apparent rotation rate of the planet in the fixed-body reference frame of the moon. Uranus rotates in a retrograde motion every 17.23 hr, and its major moons orbit in a prograde motion. The combined rotations lengthen the apparent rotation rate (synodic period) of the planet observed from each of the moons according to the approximation $T_{synodic} = 1/(1/T_{spin} - 1/T_{orbit})$, where $T_{spin}$ is the spin period of Uranus and $T_{orbit}$ is the orbital period of the moon. From the surface of the synchronously orbiting moons, the observed period of rotation is 35.1 hr for Miranda and 24.1 hr for Ariel. Each moon will experience a dominant magnetic wave that is synchronized to Uranus' observed rotation.

Figure 2 illustrates the frequency representation of Uranus' magnetic field evaluated at each moon over a 20 yr period. Table 1 illustrates the frequency and amplitude (solved using a least-squares approach on the magnetic-field time series) of the components for the set of waves modeled for each moon. The field is evaluated for each moon in its respective IAU reference frame as defined by SPICE kernels: the z axis is aligned with the spin axis (roughly aligned with spin axis of the parent planet), the y axis is in the direction of Uranus, and the x axis completes the right-handed orthogonal system, roughly in the direction of the moon's orbital motion. The frequencies of these components are associated with the synodic period of Uranus ($f_s$), the orbital frequency of the moon ($f_o$), their harmonics and the beat frequencies between these two fundamental frequencies. It is precisely these large-amplitude magnetic waves at multiple frequencies make the satellites ideal for magnetic induction investigation of their interiors. The static DC field component is also shown for completeness, but is not involved in magnetic induction.

In the Europa system, the recovery of the induction signal from two distinct orbital and synodic periods can lead to unique identification of interior properties such as ice shell thickness, ocean depth, and conductivity as they are separated in period by 10s of hours (Khurana et al. 2002). Ariel and Umbriel (second and third panels of Figure 2) do not provide the same opportunity, as the amplitude of their orbital waves are small (<<1 nT), making this approach less optimal for those objects. However, the waves associated with their higher order harmonics are moderate in amplitude (>1 nT), and thus provide information at a different frequency. Miranda on the other hand (first panel of Figure 2), is a unique case as its orbital period is similar to, but somewhat shorter than, the synodic period of Uranus due to its close proximity to the planet. Because of the similarity in frequency of the two magnetic waves, little new information can be gained from the orbital period. However, like Ariel and Umbriel, Miranda also experiences higher order harmonics that can be useful for uniquely identifying ocean properties. More on this will be provided in the discussion section. The lower beat frequency between the synodic and orbital periods of Miranda creates a long 1,044 hr wave that is enticing for its ability to probe much deeper into the interior. However, the associated complex response function at this frequency is very small, thereby inducing a signal too weak to be detected (see Section 6). For completeness, we also illustrate the magnetic waves experienced by Titania and Oberon using the field model described above. As noted above, these bodies will likely encounter a more complicated magnetic environment due to the fact that they orbit in and out of the magnetopause where associated current systems will noticeably affect the fields in which they are immersed. Because of this additional complication, we only focus on magnetic induction of the major three inner moons, Miranda, Ariel, and Umbriel.

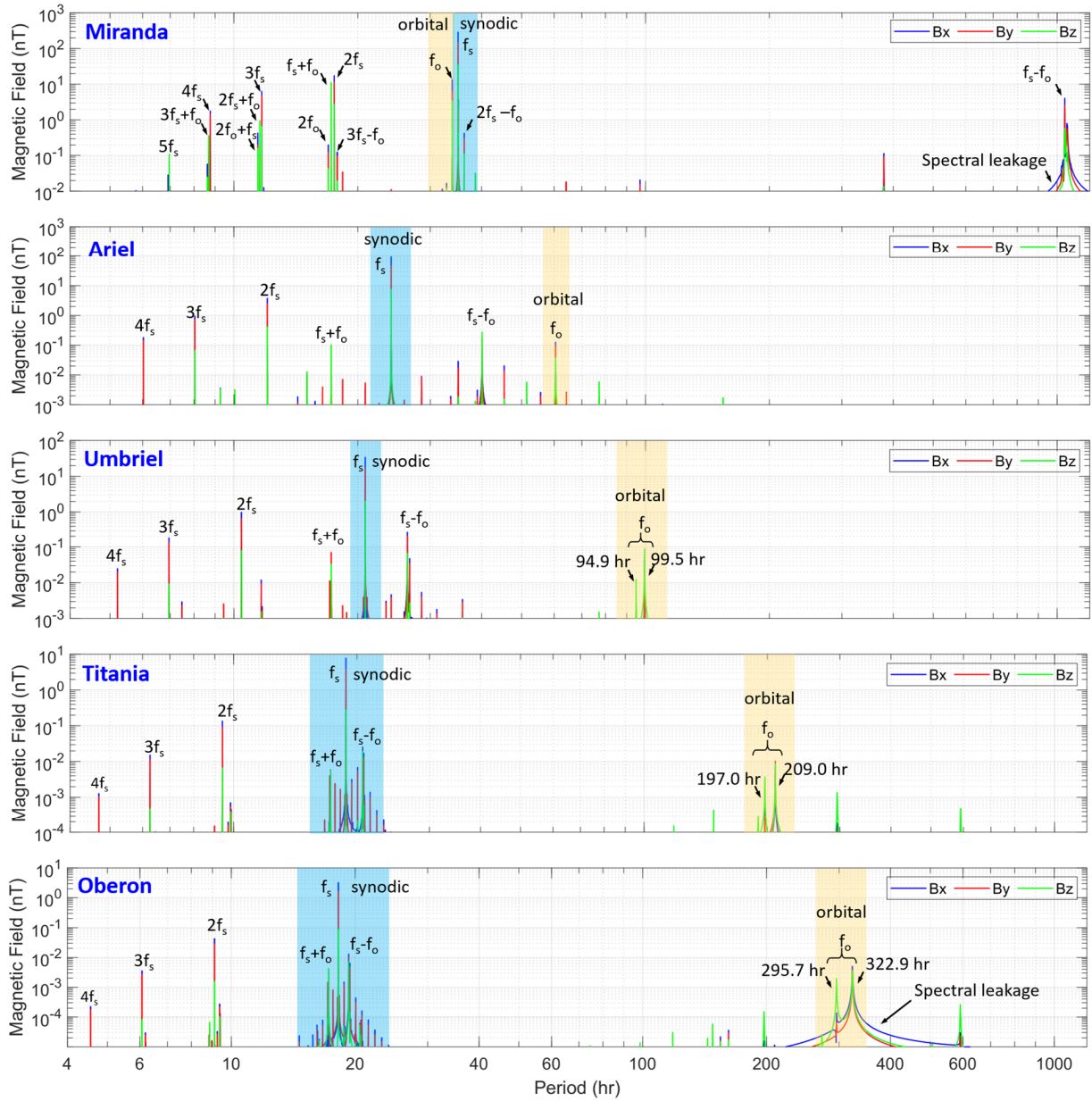

**Figure 2.** The frequency spectrum of the magnetic field at Miranda, Ariel, Umbriel, Titania, and Oberon in each's respective fixed-body frame of reference. The synodic periods are highlighted with blue and the orbital periods are highlighted in yellow. The broad frequency responses associated with the very low frequencies in the Miranda and Oberon plots are due to spectral leakage, an artifact that arises from the computation of the Fourier transform.

| | Label | Frequency (µHz) | Period (hr) | Bx (nT) | By (nT) | Bz (nT) |
|---|---|---|---|---|---|---|
| **Miranda** | DC | 0 | - | -21.29 | -0.30 | -85.97 |
| | $f_S$ | 7.93 | 35.05 | 288.33 | 144.75 | 35.01 |
| | $2f_S$ | 15.85 | 17.52 | 17.38 | 11.48 | 2.79 |
| | $3f_S$ | 23.78 | 11.69 | 6.31 | 4.73 | 0.67 |
| | $4f_S$ | 31.7 | 8.69 | 1.78 | 1.42 | 0.002 |
| | $f_O$ | 8.2 | 33.91 | 13.98 | 5.94 | 3.36 |
| | $2f_O$ | 16.4 | 16.96 | 0.21 | 0.11 | 0.089 |
| | $f_S - f_O$ | 0.27 | 1,044 | 3.32 | 2.04 | 1.09 |
| | $f_S + f_O$ | 16.12 | 17.23 | 4.19 | 0.39 | 11.33 |
| | $2f_S - f_O$ | 7.66 | 36.27 | 0.076 | 0.084 | 0.20 |
| | $2f_S + f_O$ | 24.04 | 11.55 | 0.45 | 0.13 | 0.94 |
| | $2f_O + f_S$ | 24.30 | 11.43 | 0.45 | 0.19 | 0.18 |
| | $3f_S + f_O$ | 31.96 | 8.69 | 0.14 | 0.064 | 0.37 |
| | **Label** | **Frequency (µHz)** | **Period (hr)** | **Bx (nT)** | **By (nT)** | **Bz (nT)** |
| **Ariel** | DC | 0 | - | -4.66 | -0.08 | -27.05 |
| | $f_S$ | 11.51 | 24.12 | 91.70 | 45.95 | 7.59 |
| | $2f_S$ | 23.03 | 12.06 | 3.71 | 2.46 | 0.41 |
| | $3f_S$ | 34.55 | 8.04 | 0.93 | 0.70 | 0.067 |
| | $f_O$ | 4.59 | 60.5 | 0.14 | 0.11 | 0.049 |
| | $f_S - f_O$ | 6.93 | 40.09 | 0.13 | 0.098 | 0.28 |
| | $f_S + f_O$ | 16.11 | 17.24 | 0.022 | 0.12 | 0.12 |
| | **Label** | **Frequency (µHz)** | **Period (hr)** | **Bx (nT)** | **By (nT)** | **Bz (nT)** |
| **Umbriel** | DC | 0 | - | -1.23 | -0.02 | -10.00 |
| | $f_S$ | 14.78 | 20.86 | 34.09 | 17.07 | 2.02 |
| | $2f_S$ | 26.63 | 10.43 | 0.98 | 0.65 | 0.078 |
| | $3f_S$ | 39.97 | 6.95 | 0.18 | 0.13 | 0.009 |
| | $f_O$ | 2.79 | 99.47 | 0.063 | 0.015 | 0.11 |
| | $f_S - f_O$ | 10.52 | 26.39 | 0.38 | 0.29 | 0.10 |
| | $f_S + f_O$ | 16.11 | 17.24 | 0.008 | 0.098 | 0.044 |
| | **Label** | **Frequency (µHz)** | **Period (hr)** | **Bx (nT)** | **By (nT)** | **Bz (nT)** |
| **Titania** | DC | 0 | - | -0.17 | 0.00 | -2.26 |
| | $f_S$ | 14.78 | 18.79 | 7.76 | 3.88 | 0.28 |
| | $2f_S$ | 29.55 | 9.40 | 0.136 | 0.091 | 0.007 |
| | $3f_S$ | 44.37 | 6.26 | 0.015 | 0.011 | 0.00 |
| | $f_O$ | 1.33 | 209.00 | 0.006 | 0.012 | 0.011 |
| | $f_S - f_O$ | 13.45 | 20.65 | 0.036 | 0.026 | 0.025 |
| | $f_S + f_O$ | 16.11 | 17.24 | 0.001 | 0.008 | 0.006 |
| | **Label** | **Frequency (µHz)** | **Period (hr)** | **Bx (nT)** | **By (nT)** | **Bz (nT)** |
| **Oberon** | DC | 0 | - | -0.05 | 0.00 | -0.95 |
| | $f_S$ | 15.25 | 18.21 | 3.25 | 1.63 | 0.088 |
| | $2f_S$ | 30.49 | 9.11 | 0.042 | 0.028 | 0.002 |
| | $3f_S$ | 45.76 | 6.07 | 0.004 | 0.003 | 0.000 |
| | $f_O$ | 0.86 | 322.94 | 0.009 | 0.004 | 0.003 |
| | $f_S - f_O$ | 14.39 | 19.30 | 0.014 | 0.010 | 0.011 |
| | $f_S + f_O$ | 16.11 | 17.24 | 0.001 | 0.004 | 0.008 |

**Table 1.** List of the frequencies and amplitudes of the magnetic waves encountered by each of the major moons.

## 4 Favorable Geometry for Minimal Plasma Effects

The strong tilt of Uranus' magnetic dipole axis with respect to its spin axis not only provides a sufficiently large time-varying field at each of its moons, but also provides very favorable conditions in which plasma interaction fields are minimized. As the majority of the corotating plasma is expected to be more densely confined to a disc coinciding with Uranus' magnetic equator, magnetic induction measurements would likely be optimal when the moon is farthest from this disc and the current systems associated with the flowing plasma. In the Jovian system, the moons are never very far from the plasma sheet because Jupiter's dipole axis is tilted only 9.6° with respect to its spin axis and the orbital plane of the moons. As a result, the moons spend much of their time within or near the plasma sheet (within a few $R_J$) and commonly encounter plasma induced fields that can mask the induction response of the moons (Schilling et al. 2004). This kind of masking has been observed by *Galileo* during flybys of Europa and Callisto (e.g. Zimmer et al. 2000). Additionally, the angle between Jupiter's magnetic field vector and the corotational plasma flow at the position of the moon does not deviate much from 90°, coinciding with the case where the largest perturbation to the local magnetic field occurs. This interference makes interpreting the induced field from potential oceans more complicated. This geometry is not encountered at either of the ice giants, making magnetic induction study ideal for the moons of Uranus and also Neptune's moon Triton.

Figure 3 illustrates the temporal variation of both the distance of each moon from Uranus' magnetic dipole axis (solid black line) and the angle between the magnetic field vector and the velocity vector of the plasma flow direction (dashed blue line), assumed to be corotating along the precessing magnetic equator. The plasma disc will not be perfectly confined to the magnetic equator, as centrifugal effects result from the large angle the magnetic equator makes with the spin equator. However, the magnetic equator is a good approximation for estimating the amount of time each moon spends in and around the disc. As illustrated, each of the moons spends the majority of its time away from the magnetic equator, where the plasma density is anticipated to be the largest. Miranda can be located up to 5 $R_U$ away from the magnetic equator and Oberon can be up to 20 $R_U$ away. During the few times when each of the moons resides in the region near the magnetic equator (illustrated in Figure 3 by the red horizontal bar of thickness ±1 $R_U$), the field–flow angle is at its farthest from perpendicular, a favorable geometry for being able to separate induced signals due to plasma currents. As illustrated in the figures, the field–flow angle can be ±60° away from perpendicular (corresponding to 30° and 150°) which significantly reduces the distorting effect that the plasma currents have on the local magnetic field. Note that the 60°coincides with the tilt of Uranus' magnetic axis with respect to the spin axis. This favorable geometry also occurs in the Neptune/Triton system due to the large tilt of Neptune's magnetic axis with respect to its spin axis (roughly 47°).

The low magnetospheric plasma densities anticipated in the Uranus system will further minimize plasma induced effects that may occur. Measurements by the *Voyager 2* Plasma Science experiment (PLS) indicated that plasma densities were generally below ~ a few cm$^{-3}$ (Bridge et al., 1986). These densities are very low, for example when compared to the relatively dense (~150 cm$^{-3}$) plasma that Europa encounters while within Jupiter's plasma sheet (Bagenal et al., 2015).

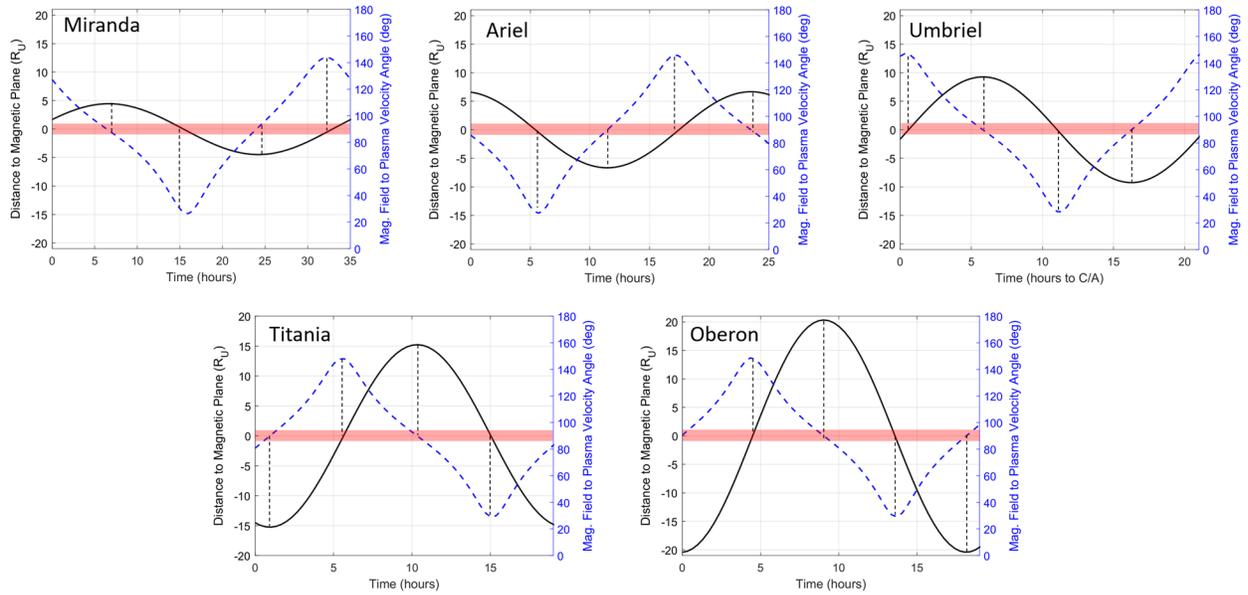

**Figure 3.** Distance from each moon to the magnetic equator (black line) of Uranus and angle between Uranus' magnetic field and corotational plasma flow direction (blue line, referred to as the field–flow angle) as a function of time for 1 full synodic period for the five major moons of the planet. The red horizontal bar indicates a ±1 $R_U$-thick region about the magnetic axis where plasma currents are expected to create the largest perturbations of the local field.

## 5 Effects of Magnetopause Currents on the Field Observed by the moons

In addition to magnetic field variability around the Uranian moons caused by the internally generated magnetic field of the planet, there are also time-dependent external sources of magnetic field that should be accounted for. Primary among these is the magnetic field arising from the electric currents that flow along the magnetopause boundary of Uranus' magnetosphere (Ness et al., 1986; Russell et al., 1989). At this boundary there is a direct interaction between the planetary magnetic field and the magnetized flow of solar wind plasma from the Sun, which flows around the Uranian magnetospheric obstacle. The magnetic field arising from magnetopause currents is present throughout the magnetosphere and surrounds all five large Uranian moons, and so magnetopause dynamics represent a further source of inducing magnetic fields at these bodies. Such contributions to an inducing field spectrum have been studied in the context of Jupiter's Galilean moons (Seufert et al., 2011) where the effects were found to be small, and have proven to be a powerful tool for probing the interior of the planet Mercury (e.g., Johnson et al., 2016).

At the large moons of Uranus, we expect the magnitude of the magnetopause-generated magnetic field contribution to be a few nT and approximately uniform throughout the magnetospheric volume, including at the location of the moons (Voigt et al., 1987). Dynamics of the magnetopause current layer will be driven by external changes in the solar wind and by the internal rotation of the planetary magnetic field. In the former case, we presently have insufficient knowledge of the state of the heliosphere at Uranus' orbit and its dependence on the solar cycle to expect any deterministic signal to result, although a signal related to solar rotation is expected (Seufert et al., 2011). In the latter case, planet-driven rotation of the magnetopause current system and the non-spherical shape of the boundary will contribute to the inducing field signals at each moon that are related to the periods of planetary rotation and the orbit of the moon of interest. Detailed magnetospheric modelling is required to resolve such effects (e.g., Voigt et al., 1987). Such modeling is beyond the scope of the present study.

## 6 Induction models of Miranda, Ariel, and Umbriel

Here, we forward model the magnetic induction response for thousands of different combinations of interior properties (ocean thickness, depth, and conductivity) and ionosphere properties (ionospheric height integrated conductivity) for the moons Miranda, Ariel, and Umbriel. Note again that we focus on the inner three major moons because they orbit well within the magnetopause boundary and are not strongly affected by these current systems. We illustrate the ability to discern certain models from others through their simulated induced magnetic moments, including distinguishing between the ocean-plus-ionosphere and ionosphere-only models in the magnetic moment space. As the moons are relatively small, gravitational effects are not sufficient to retain a significant ionosphere. For example, it has been shown that volatiles will easily escape Miranda due to its low gravity (Sori2017). However, even a low-density ionosphere might confound the induction signal from an ocean due to currents generated within this outermost conductive layer, so we take this into account in our analysis.

We use the interior modeling method described by Vance et al. (2021) to generate the complex response functions for each model. This complex function describes how the moon will respond to Uranus' time-varying magnetic field over a continuous range of frequencies. We simulate the magnetic moment $\boldsymbol{U}(t)$ using an induction model initially applied to the Jovian moons (Zimmer2000),

$$\boldsymbol{U}(t) = -\frac{4\pi}{\mu_0}\frac{R_M^3}{2}\boldsymbol{M}(t), \qquad (2)$$

where $\mu_0$ is the permeability of free space, $R_M$ is the radius of moon, and the time-varying portion of the moment vector $\boldsymbol{M}(t) = [M_x(t), M_y(t), M_z(t)]$ can be represented by the summation of scaled and phase-delayed dominant magnetic waves at various frequencies (see Figure 2),

$$\boldsymbol{M}(t) = \sum_k A_k e^{-i\phi_k}\boldsymbol{B}_{U,k}(t). \qquad (3)$$

Here, $A_k e^{-i\phi_k}$ forms the complex response function where $A_k$ is the relative amplitude and $\phi_k$ is the phase lag of the ocean's induction response at frequency $k$. The physical meaning of the amplitude and phase parameters are that they represent the relative magnitude and phase delay or lag of the induced magnetic moment with respect to that which would be induced by a perfectly conducting sphere of radius $R_M$. (For additional background, we refer the reader to the supplemental material of Vance et al. (2021), where it is shown that the complex response function $A_k e^{-i\phi_k}$ is a special case of the complex amplitude $A_n^e$ when $n = 1$, representing a uniform excitation field. Here, we negate the exponent of the complex response function, unlike the past work by Zimmer et al. (2000) and Hand and Chyba (2007), to accurately represent a positive delay in time.)

The magnetic moment induced at frequency $f_k$ is proportional to Uranus' driving field at the same frequency $\boldsymbol{B}_{U,k}(t)$, defined by

$$\boldsymbol{B}_{U,k}(t) = \boldsymbol{B}_k e^{i(2\pi f_k t + \theta_k)}, \qquad (4)$$

where $\boldsymbol{B}_k$ are the amplitudes of the Uranus' magnetic waves at frequency $f_k$ and with phase $\theta_k$, referenced with respect to the J2000 epoch. The induced dipolar magnetic field associated with the induced magnetic moment $\boldsymbol{U}(t)$ is defined by the real part of the dipole field equation,

$$\boldsymbol{B}_M(t) = Re\left\{\frac{\mu_0}{4\pi}\frac{3\big(\boldsymbol{r}\cdot\boldsymbol{U}(t)\big)\boldsymbol{r} - r^2\boldsymbol{U}(t)}{r^5}\right\}, \qquad (5)$$

where $\boldsymbol{r}$ is a spatial vector representing the point of observation with respect to the center of the moon.

We consider a multi-shell interior model for each moon, as illustrated in Figure 4. Each model includes a rocky core, hydrosphere with an ocean and overlying ice, and an ionosphere. For simplicity, the ocean and ionosphere have fixed uniform conductivity, and the core and ice are perfect insulators. The studied models consider logarithmically spaced distributions of conductivity, with varying thicknesses of the ice and ocean for each moon. Note that we don't consider the possibility of an induction field that could arise from a metallic core, as it would be located so deep below the rocky mantle that its contribution to the total induced magnetic field measured at spacecraft altitudes would be negligible. In addition, a conductive ocean would reduce the strength of the inducing field that penetrates it, further diminishing the size of any signal from a conductive core (see, e.g., Seufert et al. 2011).

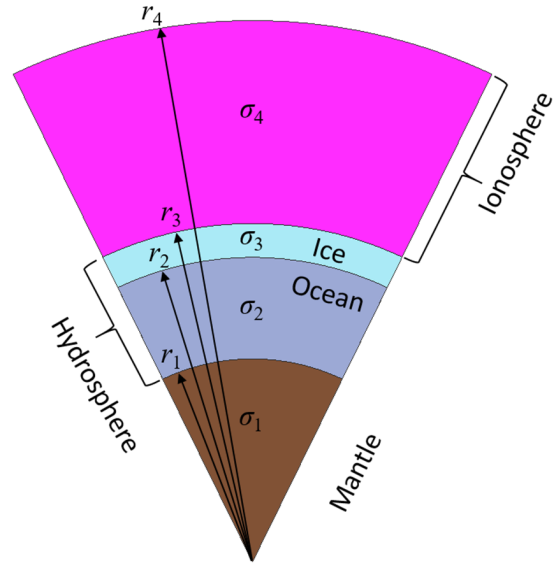

**Figure 4:** Multi-shell interior shell model used in our study. The model includes a non-conductive rocky core, a conductive ocean, a non-conductive ice shell, and uniformly conducting ionosphere. The ice shell thickness, ocean thickness, ocean conductivity, and ionospheric conductance are all adjustable parameters.

Although we do not have strong constraints regarding the possible presence of an ionosphere at any of the Uranian moons, *Voyager 2* was able to measure a highly conducting ionosphere around Triton (Majeed et al. 1990), which can be approximated by a 200-km-thick shell with a conductivity ≲0.05 S/m (Khurana et al. 2019), equivalent to an ionosphere height integrated conductivity (HIC) of up to 20 kS. We use Triton's relatively dense and conductive ionosphere to guide an upper limit for the possible contribution of ionospheres at the Uranian moons, choosing a 2.25x larger upper bound HIC of 50 kS. This conservative upper limit to ionospheric conductance serves as a stress test for the detection of possible oceans within the moons. For simplicity, we set a fixed height of the ionosphere as also done in previous studies (Hartkorn et al. 2017, Vance et al. 2020). Our modeled thickness of the ionosphere is 300 km for Ariel and Umbriel. We assume a 100 km thick ionosphere for Miranda as a worst-case ionospheric density—however, it should be noted that Miranda's small size and low gravity results in efficient escape of volatiles (Sori et al. 2017), to the extent that a substantial ionosphere at this moon is highly unlikely.

The top row of panels in Figure 5 depicts the ocean parameter modeling space used in this study for each of the moons. The size of the dot represents ocean thickness (smallest dot associated with thinnest ocean) and color represents ocean conductivity. This size and coloring code is kept consistent in in subsequent figures in order to better visualize the data in the different spaces illustrated. Forward models were generated for ocean conductivities in the range 0.1 S/m to 10 S/m in linear logarithmic increments and ocean thicknesses starting from 10 km with increments of 10 km. The seafloor depth for each moon was also varied from 16 km to 86 km at 10 km increments for Miranda, 129 km to 179 km at 10 km increments for Ariel, and 135 km to 185 km at 10 km increments for Umbriel. (The ice shell thickness is simply computed by subtracting the ocean thickness from the seafloor depth.) In addition, each combination of these models was convolved with one of 20 different ionosphere models, each with different HIC, ranging from 0 to 50 kS with step size in linear-log increments.

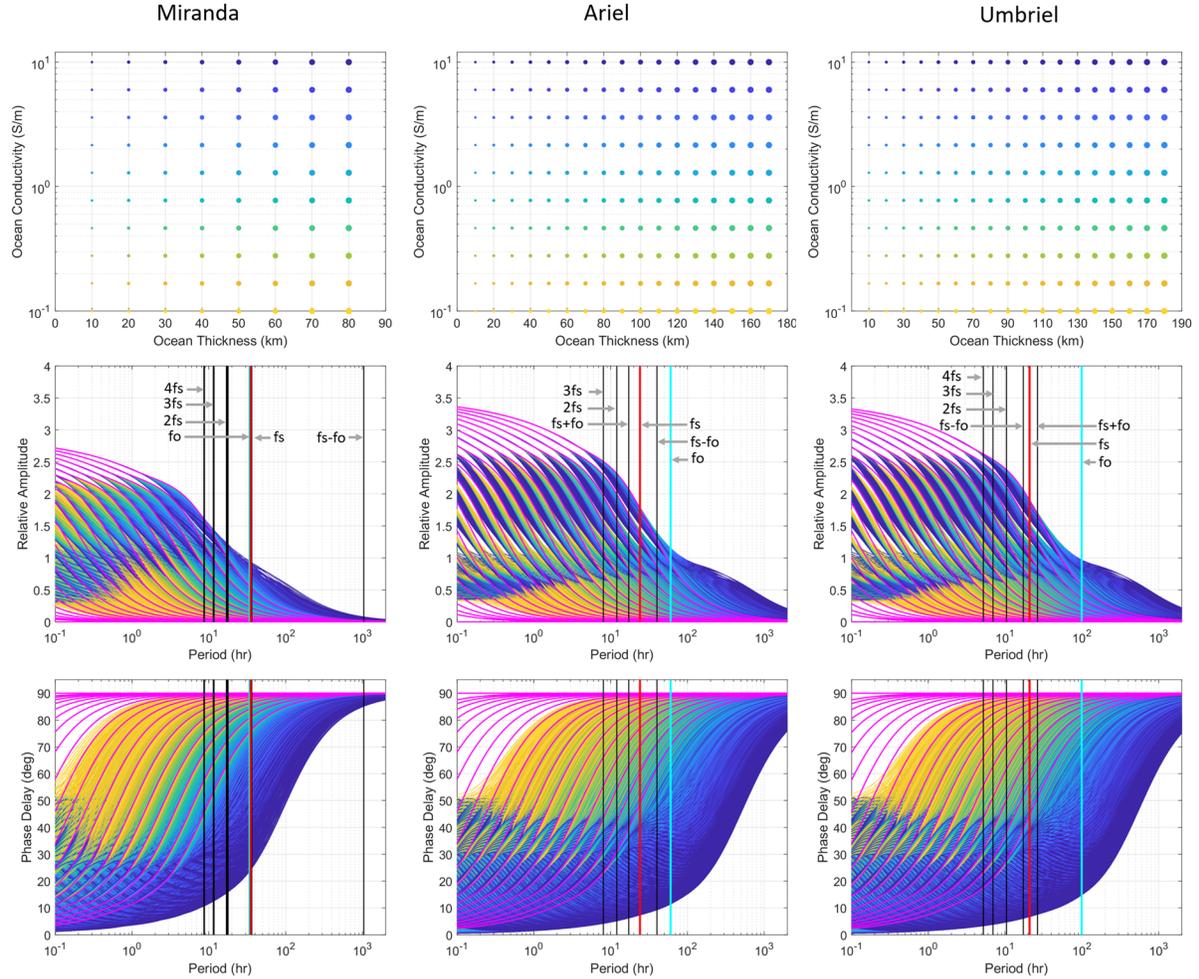

**Figure 5.** (top row) Parameter space used to forward model the various complex response functions associated with Miranda, Ariel, and Umbriel; the size of the dot indicates the ocean thickness (smallest dot represents thinnest ocean) and the color code represents ocean conductivity. (middle row) The amplitude of the complex response function for the moons of Miranda, Ariel, and Umbriel, each model colored based on the parameter space above it. Note that the complex response amplitude is normalized to the body surface, so the amplitude can be greater than 1 for highly conducting ionospheres. (bottom row) Phase delay corresponding to each amplitude above. Ionosphere-only models (e.g. no ocean conduction) are colored magenta, and are identifiable by their high amplitude at the shortest period and phase response that always asymptotes to 90°. Vertical lines represent the periods associated with the dominant magnetic waves modeled in this work. The key synodic and orbital periods are colored red and cyan, respectively. As illustrated, there are 7 periods modeled for Umbriel, 6 periods for Ariel, and 14 periods for Miranda (some are not shown, as they are too finely spaced to be viewed at this scale).

In total, 7,581 models were generated for Miranda, 18,291 models for Ariel, and 19,551 for Umbriel. The middle and bottom rows of panels in Figure 5 represent the relative amplitude $A_k$ and phase delay $\phi_k$ of the complex response function for all frequencies, for each combination of the interior and ionosphere parameters, color-coded to match the dot panels above them. The ionosphere-only models are colored magenta, and are identifiable by their high amplitude at the shortest period and phase response that always asymptotes to 90°. Also denoted on the plots by vertical lines are the periods associated with frequencies of the magnetic waves that are modeled in this work. The red vertical lines represent the synodic periods, the cyan lines represent the orbital periods, and the black lines represent the other modeled periods. In order to capture the waves with components with amplitudes of 0.1 nT or higher, a total of 14 different frequencies were modeled for Miranda, 6 for Ariel, and 7 for Umbriel.

In general, differentiation between ocean and no-ocean (magenta lines) classes is most apparent in the phase response of the complex response function. In order to better visualize the separability of these classes, as well as the different ocean classes with different interior properties, we plot the complex response function in the 2-dimensional complex plane, where the x-axis is defined by the real part of the complex response function and the y-axis is defined by the imaginary part, and the index $k$ again denotes a discrete frequency. Figure 6 illustrates the complex plane for the two key discrete frequencies for each of the moons, those being the dominant synodic and orbital periods.

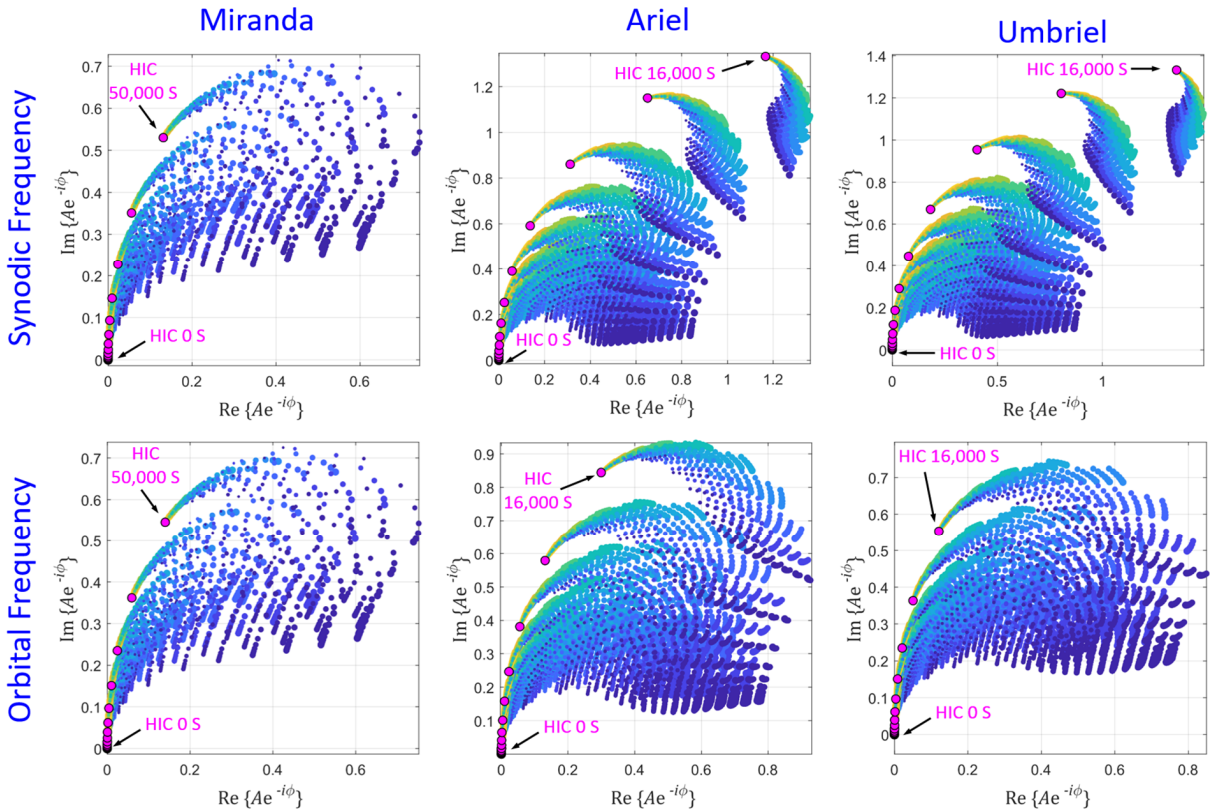

**Figure 6.** Complex plane representation of the complex response function for the studied suites of interior plus ionosphere models evaluated at the two key discrete synodic and orbital frequencies for Miranda, Ariel, and Umbriel. The coloring of the ocean models corresponds to the coloring code in Figure 5; the magenta circles correspond to ionosphere-only models. Note that the complex response amplitude is normalized to the body surface, so the amplitude can be greater than 1 for highly conducting ionospheres.

The model dot size and coloring code for Figure 6 is consistent with that illustrated in Figure 5; the magenta dots again represent the ionosphere-only models. Compared to the previous representation of the data, these plots give a better sense of the relationship between the different interior models and the ionospheres they are convolved with. For Miranda, the complex plots for the synodic period and orbital period are almost identical, simply because the periods are very similar (i.e. 34 hr vs. 35 hr). As a result, there is very little new information that can be obtained from the induction response at the orbital frequency. Additionally, even though there is a magnetic wave of appreciable amplitude with a period of roughly 1,044 hr (see Figure 2), corresponding to the lower beat of these two fundamental periods (i.e. $f_o - f_s$), the relative amplitude parameter of the complex response function at this long period is very near zero for all ocean and ionosphere models (see Figure 5) owing to the large skin depth for this slowly varying wave. This effect significantly attenuates the 1,044 hr beat frequency, rendering it useless for breaking degeneracy of the interior models. For the other two moons, the synodic and orbital periods are far enough apart in the frequency domain that new information can be obtained from the orbital response; however, because each of these moons has a nearly circular orbit with very low orbital inclination, the amplitude of the magnetic waves associated with this period are very small for both moons (e.g. the largest component, $B_x$, is less than 0.15 nT for both moons), again reducing the utility of the secondary wave to resolve the degeneracy of the interior models. Because of this low amplitude of the orbital response, the synodic period and its higher order harmonics contain most of the information and are the primary signal for distinguishing between models.

## 7 Forward Modeling the Induced Magnetic Moments

To obtain a more meaningful metric of distinguishably within the interior parameter space used in this work, we compute the total induced magnetic moment for each combination of interior and ionosphere properties (Equation 3). This allows us to work in a space with convenient units of nT and in a domain readily attainable from inverting the induced dipolar response (equation 5) that would be measured by a 3-axis magnetometer on a future spacecraft mission to Uranus.

As described previously, the induction response from each moon is dominated by each of the respective synodic waves, as the orbital waves do not carry much strength. As a result, any uncertainty in the magnetic phase of Uranus' core rotation will have a small effect on the induction response for any given orbital phase of the moon. This small effect contrasts with the environment at Triton, where the synodic (14-hr) and orbital waves (141-hr) have similar amplitudes (~ 7 nT and ~3 nT, respectively) and can have appreciable constructive or destructive interference.

Because the synodic wave is the dominant signal for probing the Uranian moons, we assess the ability to discriminate between different interior structures via their magnetic moments for two different relevant phases of this wave: those times occurring when the synodic $B_x$ component is at maximum (corresponding to minimum $B_y$) and when $B_y$ is maximum (corresponding to minimum $B_x$). This phasing and time-domain information is illustrated in Figure 7 for all three moons, where $t_1$ and $t_2$ indicate the times at which the magnetic moments are forward modeled. Also illustrated in the figure are the relatively small synodic 2nd harmonic and orbital magnetic field components, shown here for reference.

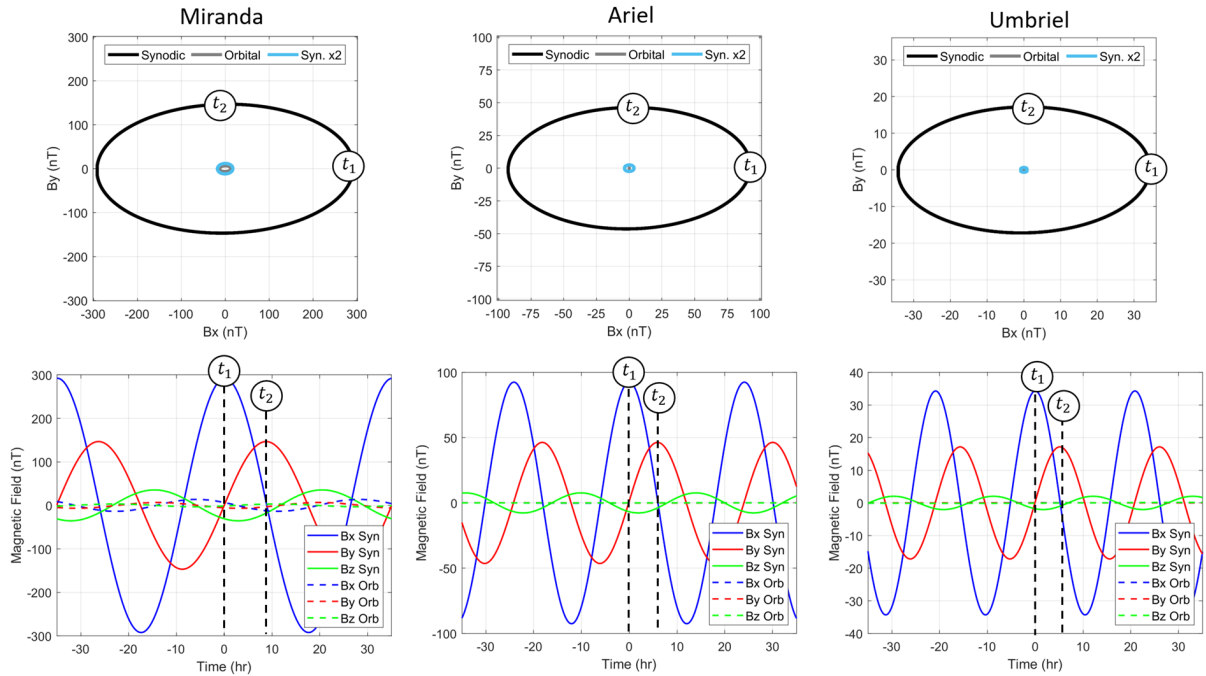

**Figure 7.** (top row) $B_x$ vs. $B_y$ components of Uranus' magnetic field at the synodic (black), synodic 2nd harmonic (cyan), and orbital (gray) periods for each of the moons. Note the insignificance of the orbital and synodic second harmonic responses compared to the dominant synodic period. The $t_1$ and $t_2$ labels indicate the two times of arrival that were simulated, corresponding to maximum $B_x$ and maximum $B_y$, respectively. (bottom row) $B_x$, $B_y$, and $B_z$ components of Uranus' magnetic field (synodic and orbital) as a function of time for the three moons.

Figure 8 illustrates the forward modeled magnetic moments for all combinations of interior and ionosphere parameters, evaluated at time $t = t_1$ when the synodic $B_x$ component is at a maximum, as defined in Figure 7. These induced magnetic moments represent a 3-dimensional space, but the $M_z(t)$ component is small compared to the other two components and therefore not shown here. It is evident in the figure that the data tend to cluster in groups with the ionosphere-only model with which they were convolved (magenta dots). Additionally, similar ocean conductivity (i.e. color) and ocean thickness (i.e. dot size) models tend to cluster within the subgroups. The thicker and highly conductive oceans are more distant from the ionosphere models they were convolved with, and the thinner and less conductive oceans lie closer, making ocean discrimination more difficult for these cases when noise sources are considered. The bottom row of Figure 8 illustrates the magnetic separation between the magnetic moments of all ocean models and the ionosphere models they were convolved with, plotted against ionosphere conductance for the three moons. The magnetic separation (MS) between the magnetic moment of a given ionosphere model $\widetilde{M}_x, \widetilde{M}_y, \widetilde{M}_z$ and magnetic moment of the $j$th ocean model it was convolved with $M_{xj}, M_{yj}, M_{zj}$ is defined by the 3-dimensional Euclidean distance between the two vectors,

$$MS_j = \sqrt{\sum_{i=x,y,z} \left( M_{ij} - \widetilde{M}_i \right)^2}. \qquad (6)$$

The increased magnetic separation between models for the moons closer to Uranus occurs because the driving field is stronger. Additionally, the thinner (smallest sized dots) and less-conductive oceans (yellow and orange shaded dots), the harder it is to distinguish from an ionosphere if both are present.

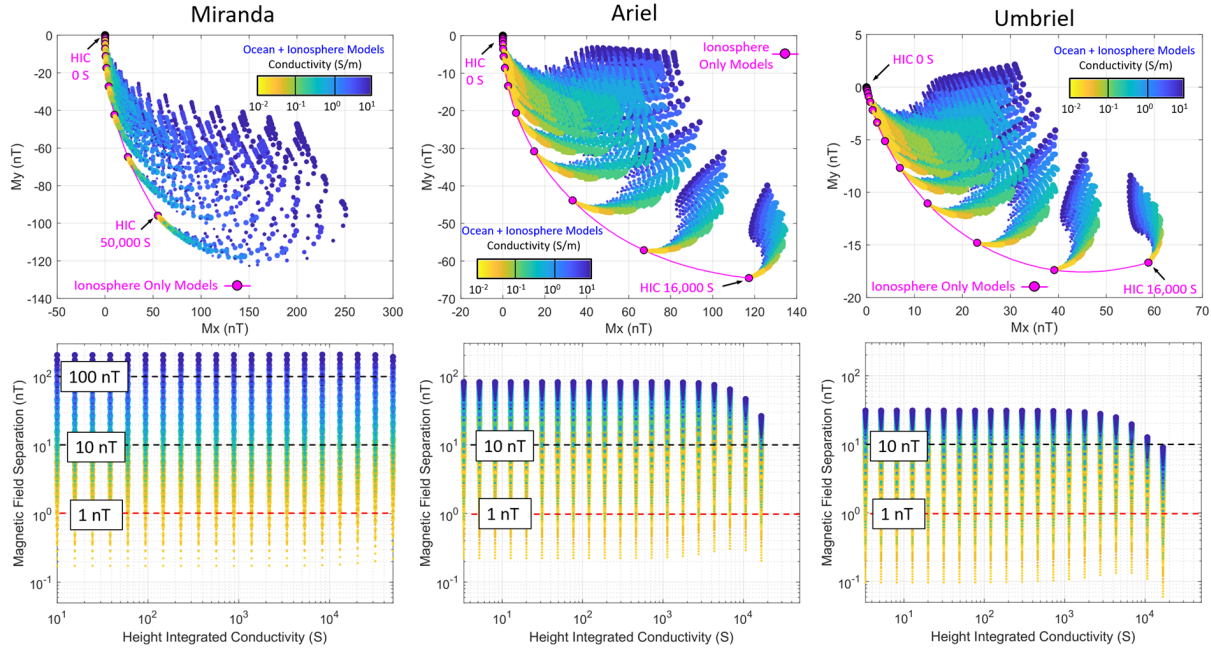

**Figure 8.** Forward modeled magnetic moment at $t = t_1$. (top row) Forward modeled magnetic moments of all combinations of interior/ionosphere models when the $B_x$ component of the synodic wave is at a maximum (correspondingly $B_y$ at a minimum) for each of the three moons studied. Color and size are again consistent with the coding in Figure 5. (bottom row) Magnetic separation between all ocean-plus-ionosphere models and ionosphere-only models with which they were convolved as a function of ionosphere HIC for each of the moons.

Figure 9 is analogous to Figure 8, but with the magnetic moments computed at time $t = t_2$ when the $B_y$ component of the synodic wave is at a maximum and the $B_x$ component of the synodic wave is zero (see Figure 7). Compared to the magnetic separation of moments in Figure 8 at $t = t_1$, there is a noticeable drop in magnetic separation in moments at $t = t_2$ for the thicker and more conductive ocean models (largest blue dots) and a small increase in the separation for the thinnest and less-conductive ocean models (smallest yellow dots). The variation in magnetic separation across a 90° phase of the synodic period is attributed to the large variation of phase delays associated with interior models simulated. Therefore, even though the driving field is maximized at $t = t_1$, the induced magnetic field will be maximized according to its associated phase delay $\phi_k$, indicating that the optimal arrival time is model dependent.

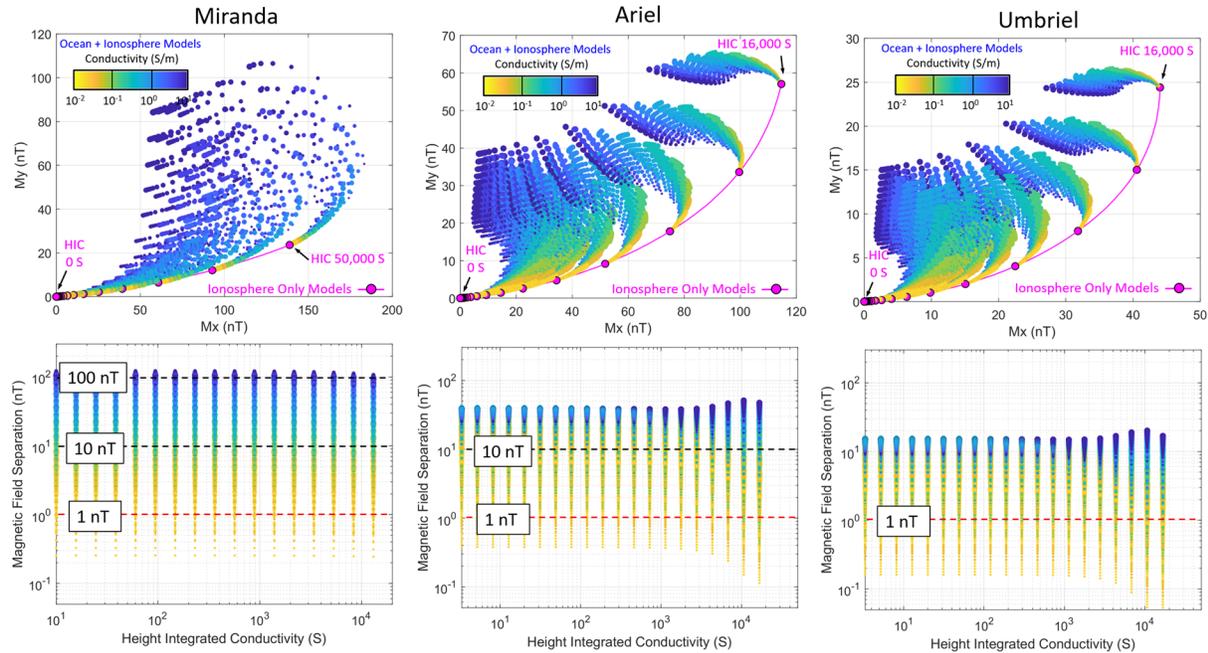

**Figure 9.** Forward modeled magnetic moment at $t = t_2$. (top row) Forward modeled magnetic moments of all combinations of interior/ionosphere models when the $B_y$ component of the synodic wave is at a maximum (correspondingly $B_x$ at a minimum) for each of the three moons studied. Color and size are again consistent with the coding in Figure 5. (bottom row) Magnetic separation between all ocean-plus-ionosphere models and ionosphere-only models with which they were convolved as a function of ionosphere HIC for each of the moons.

## 8 Discussion

In section 7, we demonstrated that it is possible to distinguish the magnetic moments associated with an ocean from those associated with an ionosphere for a wide range of ionosphere HIC. This demonstrates the feasibility of performing ocean detection from a single flyby for a wide variety of ocean scenarios if one is able to reliably invert the data for acquisition of the total magnetic moments. However, one important question remains to be addressed: What is the range of characteristics of detectable oceans when noise is considered? This question is difficult to answer without a specific planned trajectory and a Monte Carlo simulation that considers all possible noise sources (owing to plasma interaction fields, magnetic instrument noise and offsets, spacecraft position, time, attitude uncertainties, etc.). This analysis is intended for future work. However, even without such analyses, we can still get a sense of the characteristics of the oceans that are detectable by plotting the ocean-to-ionosphere magnetic field separation contours as a function of ocean thickness and conductivity, for a given ocean depth and ionosphere conductance. As shown in section 7, although the total magnetic moments can change significantly for a change in ionospheric conductance, there is not much difference in the magnetic separation of the ocean moments *relative* to moments associated with the ionospheres they were convolved with. Therefore, we illustrate a representative case to show the magnetic separation contours of models as a function of ocean thickness and conductivity for a given ocean seafloor depth and ionosphere HIC. Figure 10 illustrates these contour plots for Miranda, Ariel, and Umbriel, each at time $t_1$ and $t_2$, for assumed ionosphere HIC of ~1 kS and the deepest seafloor simulated for each body. We illustrate the results for the deepest oceans as it is expected that thicker ice shells will be present due to the very cold environment at Uranus, 19.2 AU from the Sun. These deep oceans are also the most difficult to detect, due to the $1/r^3$ falloff of the induced dipole field, and therefore represent a worst-case scenario.

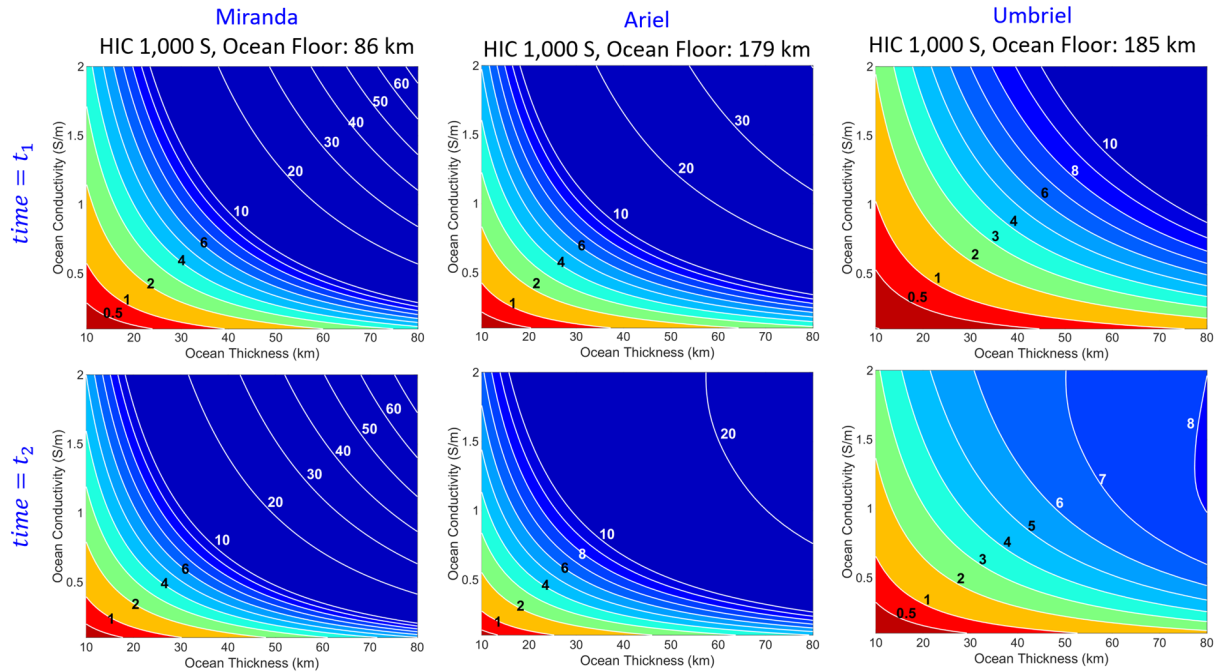

**Figure 10:** Contour plots illustrating the magnetic field separation between the total magnetic moments associated with the ocean-plus-ionosphere models and the ionosphere-only models for a representative ionosphere HIC (roughly 1,000 S) and the deepest oceans simulated for Miranda (left panels), Ariel (middle panels), and Umbriel (right panels), for two different arrival times, $t = t_1$ (top row of panels) and $t = t_2$ (bottom row of panels). The magnetic separation plotted by the contours is a proxy for the relative ease of determining whether an ionosphere alone or an ocean plus ionosphere is responsible for the hypothetical signal measured.

Again, the Miranda ocean magnetic moments exhibit the greatest separation and the Umbriel moments the least, simply due to the strength of the background magnetic field they experience due to their different proximities to Uranus. For Miranda, assuming a seafloor depth of 86 km (deepest ocean simulated) and 1 nT magnetic moment extraction noise, the ocean models that are undiscernible from ionosphere models are characterized by a parameter space bounded by a 10 km thick ocean with conductivity of 0.55 S/m and a 40 km thick ocean with conductivity of 0.1 S/m, encompassing both arrival times. Induction signals from oceans closer to the surface will be more distinguishable from the ionosphere induction signal. Assuming a 179 km deep seafloor for Ariel, the undetectable oceans within this moon span an area defined by a 10 km thick ocean with conductivity of 0.4 S/m and a 38 km thick ocean with conductivity of 0.1 S/m. Lastly, assuming a 185 km deep seafloor for Umbriel, the undetectable oceans span an area defined by a 10 km thick ocean with conductivity of 1.0 S/m and 75 km thick ocean with conductivity of 0.1 S/m. Note that at least for Miranda and Ariel, the undetectable oceans are all very likely deep and frozen out deep. Although we assert that the induction response from these oceans are not distinguishable from the induction response of an ionosphere, alternative methodologies are being developed in parallel to enhance this separation.

The contour plots of Figure 10 are not only useful for highlighting the separation of ocean moments from the ionosphere moments, but is also useful for illustrating the magnetic separation between different ocean models. This invites a second important question: How well can a detected ocean be characterized in the 3-dimensional total magnetic moment space? Figure 11 illustrates the magnetic separation contours between every ocean model and a single representative ocean model characterized with 1.3 S/m conductivity for three different ocean thicknesses for each moon at time $t = t_1$. For consistency with Figure 9, the models used to generate the contours in Figure 10 are characterized with an ionosphere HIC of ~1 kS and the deepest seafloor for each moon. Additionally, the color mapping between the two figures is identical. We simulated a 30 km, 50 km, and 70 km thick ocean for Miranda and a 30 km, 80 km, and 130 km thick ocean for Ariel and Umbriel.

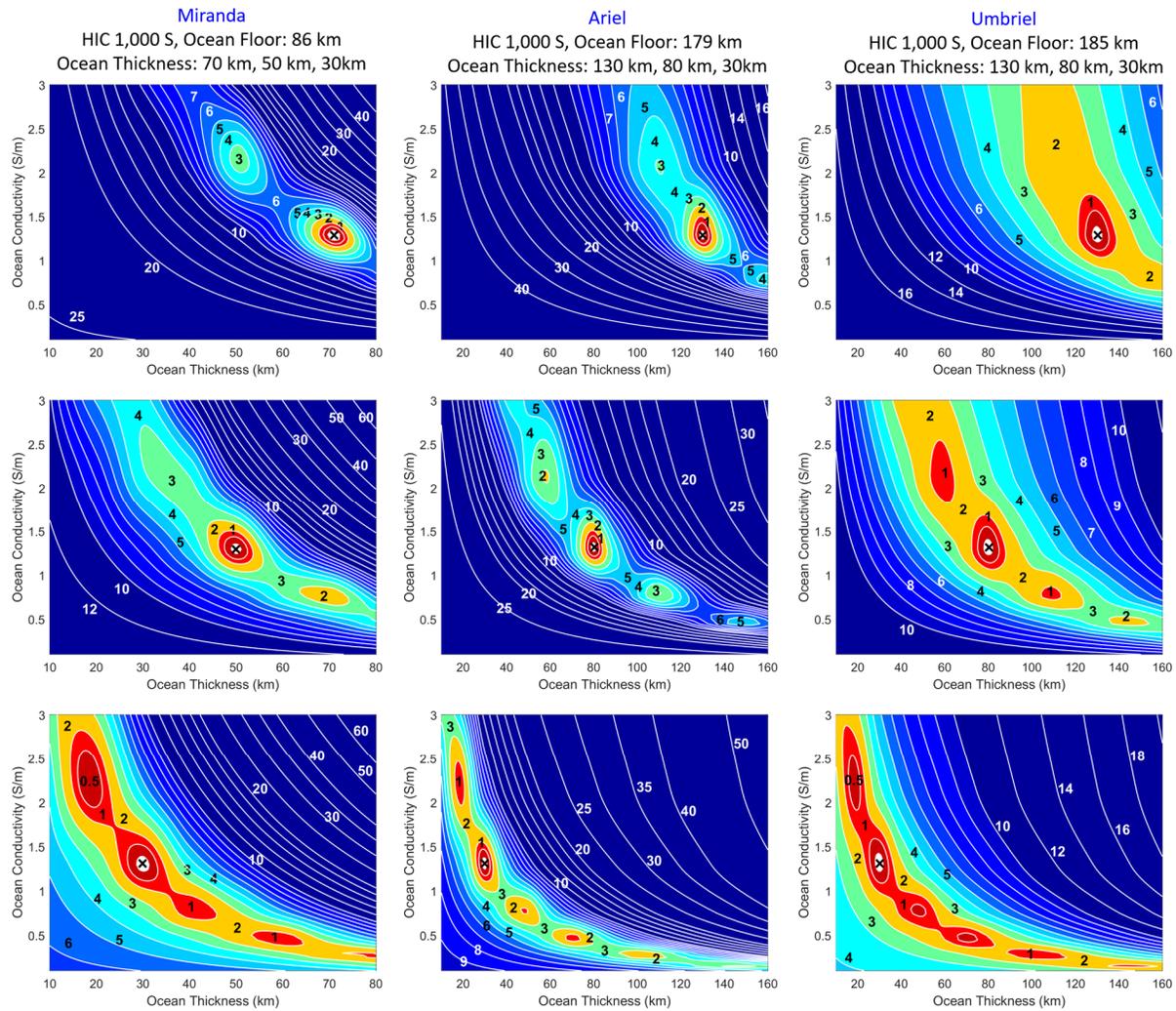

**Figure 11:** Contour plots of three representative oceans for each moon, illustrating the magnetic field separation between a single representative ocean magnetic moment, marked by the black '×', and all other ocean model moments for a given ionosphere HIC and seafloor depth. Each comparison is characterized by a band of degenerate models that would be difficult to distinguish from the example case. Outside this band (dark blue regions), the different models are expected to exhibit differences well exceeding the expected precision of measurements after accounting for noise sources.

These plots give insight into the level of noise that hinders the ability to resolve different ocean characteristics. For example, for 1 nT of noise, the oceans that fall in the space encompassed by the contours that contain the red swaths are indistinguishable as the noise has the potential to displace the model from its true 3-dimensional position in total magnetic moment space. For 2 nT of noise, the oceans that are indistinguishable reside in the contours that contain the orange swaths, and so on. It is clearly evident from the figures that the thicker and more conductive the ocean, the better the ocean resolvability. Take for example the case of Miranda with an ocean characterized by a thickness of 70 km and 1.3 S/m conductivity, marked by the '×' in the top left panel of the figure. As illustrated by the various contours surrounding this specific model, the ocean parameters can be estimated with good precision, at least here to the level of the increments used in the parameter space, that being within 10 km in thickness and within 1 S/m. Now consider the intermediate case (middle left panel), when the ocean thickness is increased to 50 km. Here, the contour areas tend to grow and no longer define unique locations, making ocean characterization more difficult when higher levels of noise are encountered (e.g. 2 and 3 nT). In the bottom left panel, the 30 km ocean thickness case is the worst scenario, as the contour regions grow in area, making characterization impossible unless the noise is kept below 0.5 nT.

Aside from increasing the number of flybys, we have also explored the possibility to further separate the models in a higher-dimensional space in order to reduce the degeneracy. Instead of using the separation distance obtained from the 3-dimensional total magnetic moment vector, one could alternatively use the separation distance obtained from the combination of the 3-element vectors for each of the individual magnetic moments (i.e. different frequencies) if extraction is possible. Mathematically, this magnetic separation is defined by

$$MS_j = \sqrt{\sum_k \sum_{i=x,y,z} \left(M_{ijk} - M_{iJk}\right)^2}, \qquad (7)$$

where $M_{ijk}$ represents the $x, y,$ or $z$ component of the $j$th ocean model at frequency $k$ and $J$ represents the index of the specific ocean model the distances are being referenced to. This separation metric is beneficial as it is not prone to destructive interference of the magnetic waves at different frequencies and therefore its capability to isolate frequencies has the ability remove degeneracy in a higher-dimensional space. Unfortunately, the Uranian moons are not optimal targets to benefit from this separation method, as the moons are dominated by a single wave (e.g. the synodic period). This implies that destructive (as well as constructive) interference of the individual waves will be minimal. We note that the degeneracy is reduced when this separation method is applied to the moons, but far from removed. Miranda, however, experiences higher-order harmonics of the synodic period that are, although small, significant enough in amplitude to slightly aid in separation of the models to reduce degeneracy. Figure 12 illustrates a representative example, which compares the magnetic separation contours for the total magnetic moments (left panel) and the individual magnetic moments (right panel), referenced to a simulated ocean with thickness of 40 km, conductivity of 1.3 S/m, seafloor depth of 56 km, and ionosphere HIC of 1 kS. Assuming a 3 nT noise limit, precisely identifying the parameter space associated with the reference ocean model in the total magnetic moment space is not possible as there are multiple islands of the 3 nT magnetic separation contour. However, when the magnetic separation is computed in a higher-dimension space, formed from the individual moments at the synodic period and its 2[nd] and 3[rd] harmonics, the degeneracy is broken and more precise identification of the ocean parameters becomes possible as only a single 3 nT island contour remains. We note that this is just one representative example in time and doesn't consider all possible phase scenarios of the waves used in the calculation. Therefore, there will be other times and or scenarios where this methodology results in either an increase or decrease in separation.

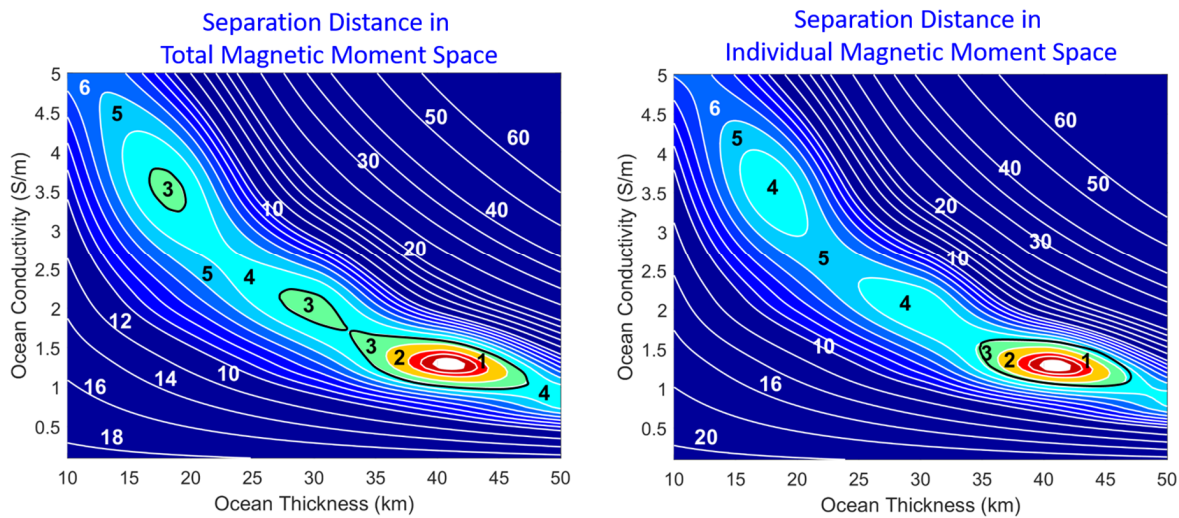

**Figure 12**: Comparison of the magnetic separation obtained from a representative ocean within Miranda (thickness of 40 km, depth of 56km, and conductivity of 1.3 S/m) with all other ocean models using (left) the total magnetic moment space and the (right) individual magnetic moment space, composed of the synodic period and its 2[nd] and 3[rd] harmonics. A modest improvement of 1-2 nT is obtained by examining the inducted magnetic moments at their individual frequencies. As illustrated, if a 3 nT total noise limit is assumed, the use of the individual magnetic moments allows more precise constrains to be placed on the ocean thickness and conductivity.

## 9 Conclusion

In this work, we have demonstrated that magnetic induction can be used to determine the existence of sub-surface oceans within the major moons of Uranus. The innermost moons of Miranda, Ariel, and Umbriel are particularly strong candidates for ocean detection using this phenomenon, as the strength and variability of the magnetic field is high at their orbital locations. For these three moons, our forward modeling indicates that ocean detection is possible for the majority of the possible interior scenarios that were considered from a single flyby, even if they possess highly conductive ionospheres. The analysis also explores the amount of noise that can be tolerated while still being able to distinguish between oceans of different depths, thicknesses, and conductivities. We demonstrate that even given realistic levels of instrument noise, these interior parameters could be constrained from a single flyby, thereby expanding the possibilities for low-cost mission architectures that seek to explore the Uranian system. Flying a heritage magnetometer with high sensitivity (e.g. $<< 100 \ pT/\sqrt{Hz}$) such as those flown on Cassini, Juno, and Magnetospheric Multiscale (MMS) missions, would be sufficient for this purpose.

## Methods

For all timing and ephemeris computation in this work, we leverage the NAIF developed SPICE toolkit for Matlab called MICE (https://naif.jpl.nasa.gov/naif/toolkit_MATLAB.html). All magnetic field models and field line visuals were implemented in MATLAB. We leverage the most recent kernel files (pck00010.tpc, ura111.bsp, gm_de431.tpc, naif0012.tls) and we use all IAU defined reference frames defined in the IAU working group report (Archinal2018).

## Acknowledgements

The research was carried out at the Jet Propulsion Laboratory, California Institute of Technology, under a contract with the National Aeronautics and Space Administration (80NM0018D0004). Funding support was provided by the JPL Strategic Research & Technology Development program for the Icy Worlds project (13-13NAI7_2-0024) and JPL FY 2019 PMCS B&P Account, Uranus Magnetosphere and Moons Investigator (UMaMI). M. Styczinski was supported by NASA Headquarters under the NASA Earth and Space Science Fellowship Program - Grant 80NSSC18K1236. A. Masters was supported by a Royal Society University Research Fellowship.

## Data Availability Statement

The panels from each figure are archived on Zenodo with DOI: 10.5281/zenodo.4750617. Each panel is stored as a Matlab .fig file where the data can be readily extracted. Currently, the data is restricted access until publication. Until then, only access requests by journal reviewers and editors will be given.